\documentclass[11pt]{article}

\usepackage[final]{acl}

\usepackage{times}
\usepackage{latexsym}
\usepackage[T1]{fontenc}
\usepackage[utf8]{inputenc}
\usepackage{microtype}
\usepackage{inconsolata}
\usepackage{graphicx}
\usepackage{booktabs}
\usepackage{multirow}
\usepackage{xcolor}
\usepackage{colortbl}
\usepackage{amsmath,amssymb}
\usepackage{array}
\usepackage{booktabs}
\usepackage{makecell}

\usepackage{subcaption}
\usepackage[most]{tcolorbox}
\usepackage{appendix}

\definecolor{oursrow}{RGB}{230,241,251}   % light blue: our model rows
\newcommand{\ours}[1]{\rowcolor{oursrow}#1}

\title{Argus-Retriever: Vision-LLM Late-Interaction Retrieval with Region-Aware Query-Conditioned MoE for Visual Document Retrieval}

\author{
  \textbf{Abdelrahman Abdallah\textsuperscript{1}, Mahmoud Abdalla\textsuperscript{2}, Mohammed Ali\textsuperscript{1}, Adam Jatowt\textsuperscript{1}} \\
  \textsuperscript{1}University of Innsbruck \quad \textsuperscript{2}Chungbuk National University \\
  \texttt{\{abdelrahman.abdallah,adam.jatowt\}@uibk.ac.at}
}

\begin{document}
\maketitle

\begin{abstract}
Late-interaction vision-language retrievers represent each
document page as many visual token embeddings and score queries
with MaxSim. In systems such as ColPali, ColQwen, ColNomic, and
Nemotron ColEmbed, the document embeddings are produced without
seeing the query, so the same page is represented identically for
a table lookup, a chart question, and a layout-sensitive
evidence request. We introduce \textbf{Argus}, a family of
query-conditioned late-interaction retrievers built on
Qwen3.5-VL. Argus adds a region-aware Mixture-of-Experts module:
the query encoder produces both retrieval embeddings and a
compact context vector, the document page is pooled into
spatial regions, and a query-aware router selects latent experts
per region before MaxSim. The output remains a multi-vector
index compatible with ColPali-style retrieval, but the document
representation is now dependent on the query (i.e., $\mathbf{D}(q)$). All Argus models use a
1024-dimensional retrieval head, compared with the
2560-dimensional and 4096-dimensional heads of recent
state-of-the-art systems, and are trained on roughly 9\% of the
available public supervision rather than the full pool. The 9B
model reaches \textbf{92.67} NDCG@5 on ViDoRe V1 and
\textbf{86.0} NDCG@5 on the combined V1+V2 leaderboard, the
highest reported value for an open late-interaction model on
the combined leaderboard. Wrapped in a Qwen3.6-27B
agentic retrieval pipeline on ViDoRe V3, Argus-9B further
improves its NDCG@10 from 60.28 to \textbf{64.80} over public tasks, showing that the same retriever serves both
as a strong standalone system and as a search primitive for
iterative LLM agents.\footnote{The code and data are available: \url{https://github.com/DataScienceUIBK/Argus-Retriever}}
\end{abstract}

\section{Introduction}
\begin{figure}[t]
  \centering
   \includegraphics[width=\linewidth]{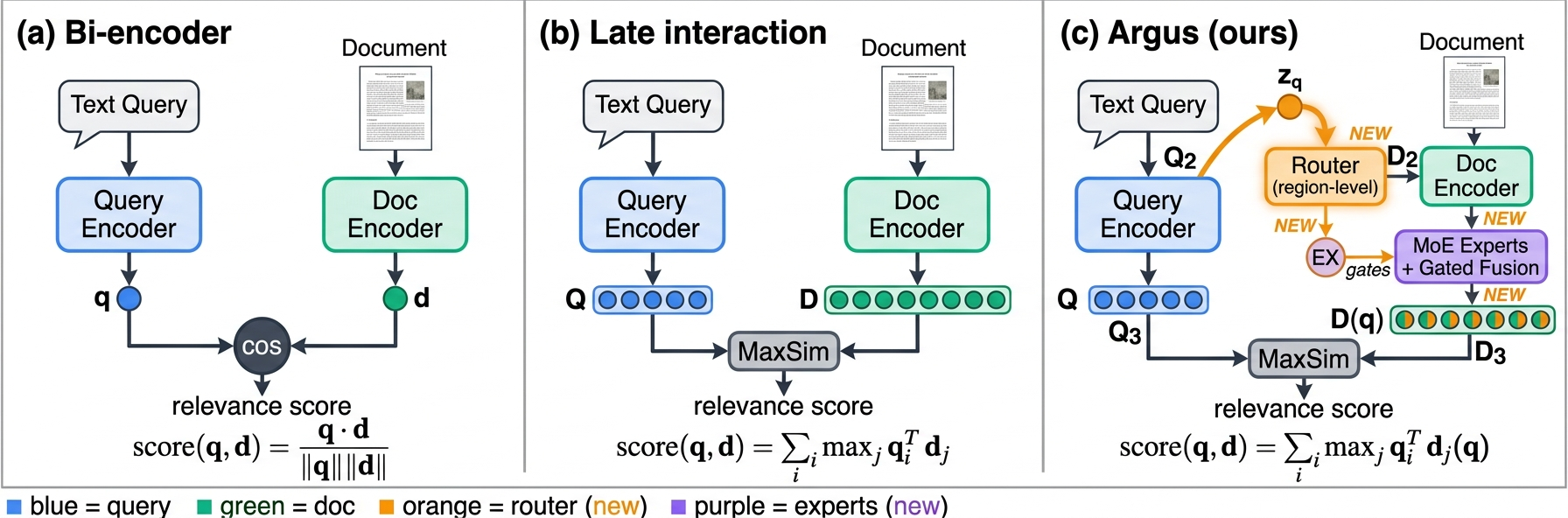}

  \caption{\textbf{Where does the query interact with the
  document?} Bi-encoders compare single vectors, while standard
  late-interaction models compare query tokens against fixed
  document tokens. Argus keeps MaxSim scoring, but adds a
  query-conditioned region router and MoE expert fusion.} 

  \label{fig:teaser}
\end{figure}
Retrieval-Augmented Generation (RAG) has become a standard way
to ground language models in external knowledge
\citep{lewis2020rag,abdallah2025dear}. The first
step in this pipeline is retrieval: given a user query, the
system must find the pages or passages that contain the
evidence needed by the generator. Many strong retrieval models
have been developed for clean text
\citep{karpukhin2020dpr,wu2024stark,abdallah2026llm,zhan2021optimizing}, but practical
enterprise and scientific corpora often arrive as PDFs,
presentation slides, scanned pages, financial reports, and
technical manuals. In these documents, the answer may be stored
in a table cell, a chart axis, a figure caption, or a formula.
Pure text extraction can miss this information, and OCR-based
pipelines often ignore the layout that makes the evidence
interpretable \citep{huang2022layoutlmv3}.

Visual Document Retrieval (VDR) addresses this problem by
retrieving rendered document pages directly from their images.
Recent VLMs can recognize rendered text,
layout, and visual structure from pixels, which has made them
natural backbones for VDR \citep{faysse2024colpali,
moreira2026nemotron}. The strongest VDR systems usually combine
a vision-language backbone with the late-interaction mechanism
introduced by ColBERT \citep{khattab2020colbert}. Instead of
compressing a page into a single vector, late-interaction
models project a document image into many patch-level
embeddings. A query is also represented as multiple token
embeddings, and relevance is computed with MaxSim, which lets
each query token match its best document token. This design
preserves fine-grained evidence that bi-encoder pooling tends
to discard.

Despite this progress, late-interaction VDR retrievers compute
document embeddings without seeing the query
\citep{faysse2024colpali, colqwen2card, nomic2024multimodal,
moreira2026nemotron}. Argus removes this restriction by adding
a query-conditioned region router and a Mixture-of-Experts
fusion inside the document encoder, so the final document
representation becomes $\mathbf{D}(q)$ while remaining a
multi-vector index compatible with MaxSim scoring
(Figure~\ref{fig:teaser}; details in \S\ref{sec:argus}). This
design is meant to recover part of the adaptivity of
cross-encoders \citep{nogueira2019monobert} while preserving
the practical benefits of late interaction. Argus also targets a practical training and storage regime.
Rather than scaling the retrieval head to
several thousand dimensions or relying on model soup and post-training checkpoint
merging \citep{wortsman2022modelsoups}, all Argus models use a 1024-dimensional
late-interaction head and are trained from a balanced subset of
the available public supervision. This gives the paper a second
axis beyond raw leaderboard accuracy: Argus improves the
accuracy and adaptivity of visual document retrieval while
keeping the embedding width and supervised training mixture
modest.

\paragraph{Contributions.}                                                             \textbf{(1) }We introduce \textbf{Argus}, a query-conditioned                            late-interaction VLM retriever. A region-aware router selects                        
latent experts that produce a query-dependent multi-vector
document representation $\mathbf{D}(q)$, scored against the
query tokens by MaxSim.
\textbf{(2) }The router conditions on three signals — per-region
content, 2D coordinates, and a pooled query context
$\mathbf{z}_q$ — so the same page routes differently for
different queries while the output stays a ColPali-compatible
multi-vector index.
\textbf{(3) }The design is deployable: the image encoder runs once
per page offline; only the query encoder, router, gated
fusion, projection head, and MaxSim run at query time against
cached visual grids.

\section{Related Work}

\subsection{Visual Document Retrieval}

Early multimodal retrievers followed the Contrastive Language–Image Pre-training (CLIP) recipe, encoding
an image and a text query into global vectors and comparing
them with a single similarity score \citep{radford2021clip,abdallah2026mm}.
That design is simple and fast, but it is a poor fit for
pages whose evidence is spread across table cells, captions,
figures, formulas, and layout. Visual document retrieval moved
the field closer to the document itself by retrieving rendered
page images instead of relying only on OCR text. ColPali
\citep{faysse2024colpali} was the key step in this direction:
it used a vision-language backbone to turn a page image into
many visual token embeddings and scored them with late
interaction. The same recipe has since been extended with
stronger backbones and training recipes, including ColQwen,
ColNomic \citep{nomic2024multimodal}, and Nemotron ColEmbed
\citep{moreira2026nemotron}. Modern VLM backbones make this possible 
because they already carry strong document understanding 
capabilities. PaliGemma\citep{paligemma}, Qwen2-VL \citep{qwen2vl}, Qwen3-VL
\citep{qwen3vl}, and NVIDIA Eagle 2 \citep{eagle2} can read
rendered text and use layout cues directly from pixels. Argus
builds on this visual retrieval line.

\begin{figure*}[t]
\centering
\includegraphics[width=0.9\linewidth]{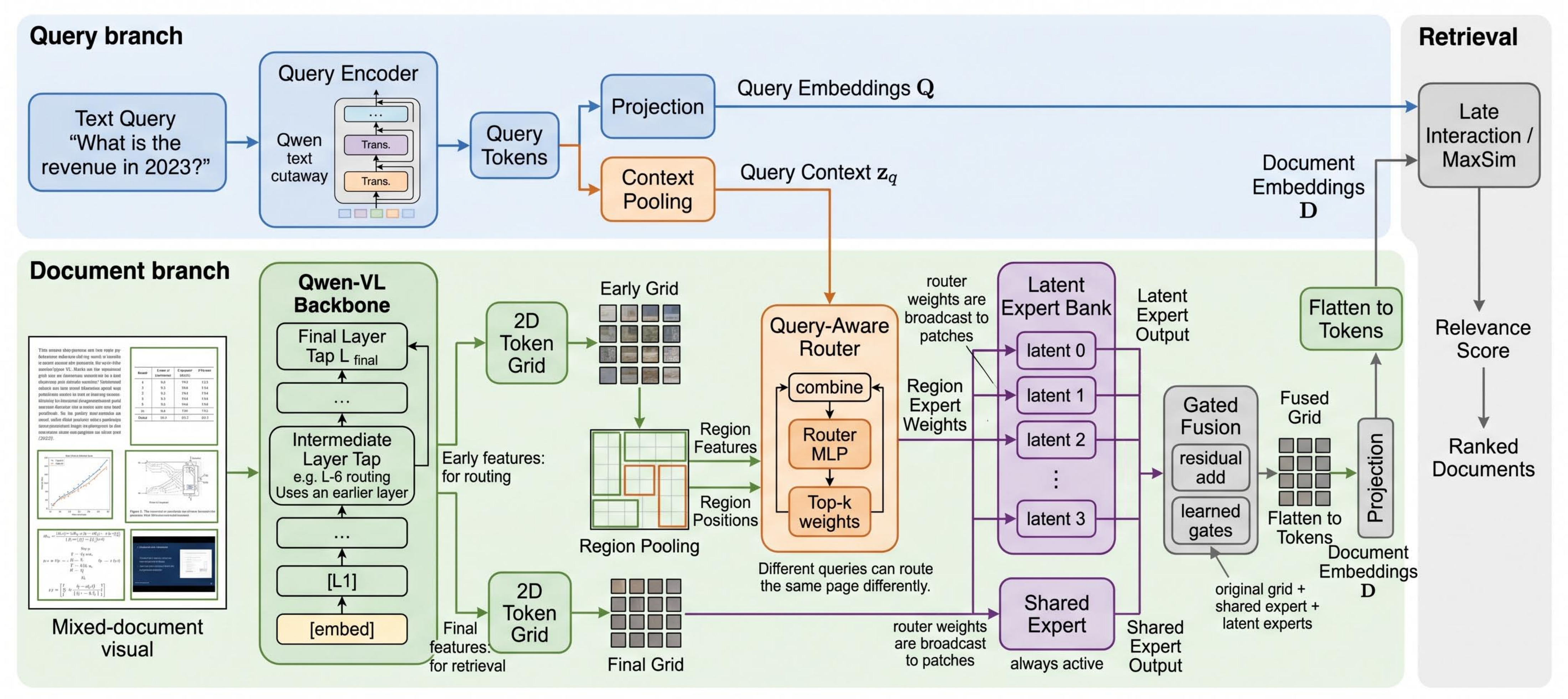}

\caption{\textbf{Argus architecture.} The query branch yields
retrieval embeddings $\mathbf{Q}$ and a pooled context
$\mathbf{z}_q$ for routing. The document branch taps Qwen3.5-VL
at two depths: $L{-}5$ feeds region pooling and the
query-aware router; $L$ feeds the latent expert bank and shared
expert. Top-$k$ router weights are broadcast to patches to mix
expert outputs, and a gated residual fusion produces the
query-conditioned grid $\mathbf{D}(q)$, scored against
$\mathbf{Q}$ by MaxSim. At inference, the image encoder runs
once per page; cached grids and region features are reused
across queries (Appendix~\ref{sec:appendix-inference}).}
\label{fig:main}
\end{figure*}

\subsection{Late Interaction}

The MaxSim operator \citep{khattab2020colbert} defines
relevance between a query $q = (q_1, \dots, q_m)$ and a
document $d = (d_1, \dots, d_n)$ as
\begin{equation}
\label{eq:maxsim}
\text{MaxSim}(q,d) \;=\; \sum_{i=1}^{m} \max_{j=1,\dots,n}\;
q_i^\top d_j .
\end{equation}

For each query token, MaxSim selects its best-matching
document token, and the resulting per-token scores are summed
into a single relevance score. We use this scoring function
throughout. Computing MaxSim for a single (query, document)
pair requires $m \times n$ inner products, which reduce to a
single matrix multiplication on GPU; the dominant cost of
late interaction is therefore the storage of all per-token
document embeddings, which we revisit in
\S\ref{sec:deployment}.

\subsection{Contrastive Learning}
\label{sec:contrastive-bg}

Late-interaction retrievers are trained with the contrastive
InfoNCE objective \citep{infonce}:
\begin{align}
\mathcal{L}_{\text{NCE}}
&= -\log
\frac{\exp(\text{sim}(q,d^+)/\tau)}
{\sum_{d_i \in \{d^+\} \cup D_N}
\exp(\text{sim}(q,d_i)/\tau)} ,
\end{align}
where $\text{sim}(\cdot,\cdot)$ is MaxSim, $\tau$ is a
temperature, and $D_N$ is the negative set. The Argus training
objective uses this loss with in-batch negatives only, as
detailed in \S\ref{sec:training-objective}.

\subsection{Mixture of Experts in Retrieval}
\label{sec:moe-bg}

Mixture-of-Experts layers use a router to send each input to a
small subset of specialist networks \citep{shazeer2017outrageously,
switch}. This idea is common in language models because it
increases capacity without activating every parameter on every
example. In retrieval the idea has been explored more sparingly:
\citet{cai2023came} apply a competitively-learned MoE to
first-stage text retrieval; \citet{nussbaum2025moeembed} train
sparse-MoE text embedding models; and \citet{sokli2025moeretrieval}
study MoE in dense retrieval. To our knowledge none of these
prior systems makes the routing decision depend on the
\emph{query} at the document side, and none target visual
document retrieval, where the main pressure is not only model
capacity but also indexing and query-time cost.

\section{Argus: A Family of Region-Aware MoE Late-Interaction Models}
\label{sec:argus}

\subsection{Model Family}

Argus is a family of three late-interaction VLM retrievers, all
built on Qwen3.5-VL backbones, all using MaxSim scoring at
embedding dimension $d=1024$ (Table~\ref{tab:family}). The
three sizes share the same query-conditioned region-aware MoE
module described in \S\ref{sec:argus-module}; only the backbone
differs. 

\begin{table}[t]
\centering
\small
\setlength{\tabcolsep}{4pt}

\caption{The Argus family. All sizes share the same
query-conditioned MoE ($K{=}4$ latent + 1 shared expert,
top-$2$ routing, region $r{=}4$, router tap $L{-}5$) and a
1024-dim retrieval head.}
\label{tab:family}
\begin{tabular}{lrcc}
\toprule
\textbf{Model} & \textbf{\#Params} & \textbf{Dim} & \textbf{Experts} \\
\midrule
\textit{Argus-2B} & 2.32\,B & 1024 & 4 (top-2) \\
\textit{Argus-4B} & 4.71\,B & 1024 & 4 (top-2) \\
\textit{Argus-9B} & 8.82\,B & 1024 & 4 (top-2) \\
\bottomrule
\end{tabular}
\end{table}

\subsection{Architecture Overview}
\label{sec:argus-module}

Figure~\ref{fig:main} shows the full training-time architecture.
Argus reads a text query and a document image through a shared
Qwen3.5-VL backbone, taps the backbone at two depths to obtain
a routing-friendly intermediate visual grid and a
retrieval-friendly final visual grid, and inserts a
query-conditioned region-aware MoE module between the final
layer and the projection head. Figure~\ref{fig:main} shows the
training-time forward pass; at inference, the document side is
cached once per page and only the query-side path runs per
query, so the query-conditioned design remains compatible with
offline indexing (Appendix~\ref{sec:appendix-inference}).

\subsection{Query Branch: $\mathbf{Q}$ and $\mathbf{z}_q$}

The text query is encoded by the shared backbone and projected
to retrieval dimension $d{=}1024$ to give per-token embeddings
$\mathbf{Q} \in \mathbb{R}^{m \times d}$ used in MaxSim. We
additionally pool $\mathbf{Q}$ with the attention mask and L2
normalize to obtain a single query context vector
\begin{equation}
\mathbf{z}_q \;=\; \frac{\sum_{i} a_i \, \mathbf{Q}_i}
{\bigl\| \sum_{i} a_i \, \mathbf{Q}_i \bigr\|_2}
\;\in\; \mathbb{R}^d ,
\end{equation}
where $a_i \in \{0,1\}$ is the query attention mask.
$\mathbf{z}_q$ is \emph{not} used directly in MaxSim. It is
sent into the document branch to condition the router, so the
same $\mathbf{Q}$ that scores documents also shapes how those
documents are encoded.

\subsection{Document Branch: Region Pooling at Two Depths}

We extract two hidden states from the document forward pass:
$\mathbf{H}_{\text{router}} = \mathbf{H}^{(L-5)}$ for routing
and $\mathbf{H}_{\text{final}} = \mathbf{H}^{(L)}$ for
retrieval. The intermediate tap preserves layout and visual
structure useful for region-level decisions; the final tap
carries the strongest semantic features for retrieval.
This separation is consistent with probing studies showing
that structural and positional information peaks in
middle-to-late transformer layers, before the final layers
specialize for the task head
\citep{tenney2019bert,karamcheti2024prismatic}, and with the
standard VLM practice of taking vision features from a layer
\emph{before} the final one \citep{liu2023llava}.
After
selecting image-token positions, we reshape both to 2D grids
\begin{equation}
\mathbf{G}_{\text{early}}, \;\mathbf{G}_{\text{final}} \;\in\;
\mathbb{R}^{H \times W \times h} .
\end{equation}
We pad each grid to a multiple of the region size $r{=}4$ and
average-pool valid patches within each $r \times r$ block to
obtain $N$ region tokens $\mathbf{R} \in \mathbb{R}^{N \times h}$
from $\mathbf{G}_{\text{early}}$, a region validity mask, and
normalized box coordinates $\mathbf{c} \in \mathbb{R}^{N \times 4}$
of the form $[x_0, y_0, x_1, y_1]$ for each region.
\begin{table*}[t]
\centering

\setlength{\tabcolsep}{2.5pt}
\caption{Retrieval accuracy (NDCG@5) on the ViDoRe V1+V2
leaderboard. Argus rows are shaded; per-column best values
across listed models are bolded. Baselines listed in
\S\ref{sec:baselines}.}
\label{tab:v12}
\resizebox{0.9\textwidth}{!}{%
\begin{tabular}{l*{10}{c}*{4}{c}|ccc}
\toprule
& \multicolumn{10}{c}{\textbf{ViDoRe V1}} &
\multicolumn{4}{c|}{\textbf{ViDoRe V2}} &
\multicolumn{3}{c}{\textbf{Average}} \\
\cmidrule(lr){2-11} \cmidrule(lr){12-15}
\textbf{Model} & ArXiv & DocVQA & InfoVQA & Shift & AI & Ene & Gov
& Hlt & TabF & TatD & BMed & ESG & ESGHL & Econ & V1 & V2 & Avg \\
\midrule
ColPali v1.3            & 83.3 & 58.4 & 85.5 & 77.3 & 97.4 & 94.6 & 96.1 & 97.4 & 86.7 & 70.8 & 54.6 & 54.9 & 58.5 & 48.6 & 84.8 & 54.2 & 76.0 \\
ColSmol-500M            & 74.9 & 58.0 & 87.2 & 66.8 & 98.4 & 94.6 & 94.7 & 98.0 & 74.7 & 77.5 & 44.9 & 40.4 & 52.4 & 36.4 & 82.5 & 43.5 & 71.4 \\
ColQwen2 v1.0           & 88.0 & 61.5 & 92.5 & 89.9 & 99.0 & 95.9 & 95.5 & 98.8 & 89.0 & 82.2 & 56.3 & 52.5 & 60.4 & 50.6 & 89.2 & 55.0 & 79.4 \\
ColQwen2.5              & 88.9 & 63.6 & 92.5 & 87.9 & 99.6 & 96.1 & 95.8 & 98.0 & 90.8 & 82.1 & 59.2 & 58.3 & 66.4 & 53.3 & 89.5 & 59.3 & 80.9 \\
ColNomic-3b             & 88.1 & 61.3 & 92.8 & 90.1 & 96.3 & 97.4 & 96.6 & 98.3 & 94.4 & 83.2 & 62.5 & 49.2 & 57.0 & 53.3 & 89.9 & 55.5 & 80.0 \\
ColNomic-7b             & 88.3 & 60.1 & 92.2 & 89.3 & 98.8 & 96.3 & 95.9 & 99.3 & 96.0 & 81.1 & 63.4 & 54.8 & 68.7 & 56.2 & 89.7 & 60.8 & 81.5 \\
Sauerkraut-4b           & 91.8 & 67.0 & 94.2 & 90.5 & 99.6 & 96.5 & 96.2 & 100.0 & 89.5 & 82.7 & 60.0 & 58.8 & 69.2 & 56.0 & 90.8 & 61.0 & 82.3 \\
Sauerkraut-8b           & 93.8 & 64.7 & 94.5 & 90.4 & 98.6 & 96.5 & 96.8 & 99.3 & 92.2 & 84.0 & 64.9 & 58.4 & 70.8 & 57.6 & 91.1 & 62.9 & 83.0 \\
Nemotron-colembed-3b-v1 & 88.4 & 66.2 & 94.9 & 90.7 & 99.6 & 96.6 & 97.8 & 99.3 & 95.9 & 80.6 & 64.3 & 58.6 & 75.4 & 55.9 & 91.0 & 63.5 & 83.2 \\
Nemotron-colembed-4b-v2 & 92.0 & 67.4 & 93.3 & 92.3 & 99.3 & 96.2 & 98.0 & 98.5 & 98.1 & 81.2 & 64.0 & 62.4 & 71.4 & 57.7 & 91.6 & 63.9 & 83.7 \\
Ops-Colqwen3-4B         & 91.8 & 66.5 & 94.0 & 90.8 & 99.6 & 97.3 & 98.0 & 99.6 & 93.6 & 82.4 & 64.9 & 65.6 & 78.6 & 61.9 & 91.4 & 67.8 & 84.6 \\
Nemotron-colembed-8b-v2 & 93.1 & 68.1 & 94.6 & 93.3 & 100.0 & 97.9 & 98.9 & 99.6 & 97.7 & 83.4 & 67.0 & 59.0 & 73.2 & 60.4 & 92.7 & 64.9 & 84.7 \\
\midrule
\ours{}\textit{Argus-2B} & 90.3 & 67.5 & 95.0 & 91.3 & 99.6 & 97.3 & 97.3 & 99.3 & 93.4 & 84.0 & 65.0 & 59.9 & 69.4 & 51.9 & 91.5 & 61.5 & 82.9 \\
\ours{}\textit{Argus-4B} & 91.0 & 67.7 & 94.6 & \textbf{94.7} & 99.6 & \textbf{97.9} & 97.8 & 99.6 & 95.3 & 84.8 & 64.4 & 62.2 & 69.9 & 59.8 & 92.3 & 64.1 & 84.2 \\
\ours{}\textit{Argus-9B} & 92.3 & 68.1 & 94.3 & 93.6 & 99.6 & 97.3 &\textbf{ 98.9} & 99.6 & 97.5 & \textbf{85.5} & 66.2 & \textbf{67.6} & \textbf{79.1} & \textbf{63.8} & \textbf{92.7} & \textbf{69.2} & \textbf{86.0} \\
\bottomrule
\end{tabular}
}
\end{table*}
\subsection{Query-Aware Router}

For each region $r$, the router input is the sum of three
projected signals: region content, region position, and query
context, all lifted to the backbone hidden dimension $h$:
\begin{equation}
\mathbf{u}_r \;=\; \mathbf{R}_r \;+\; \mathbf{W}_{\text{coord}}
\mathbf{c}_r \;+\; \mathbf{W}_{\text{query}} \mathbf{z}_q .
\end{equation}
where $\mathbf{W}_{\text{coord}} \in \mathbb{R}^{h \times 4}$
and $\mathbf{W}_{\text{query}} \in \mathbb{R}^{h \times d}$ are
learned linear projections (no bias) that lift the
4-dimensional region coordinates $\mathbf{c}_r$ and the
$d$-dimensional pooled query context $\mathbf{z}_q$ into the
backbone hidden dimension $h$.
Addition (rather than concatenation) keeps the router input
dimension fixed at $h$ and acts as a positional and query bias
on the region content, in the same spirit as additive
positional embeddings in Transformers. A two-layer MLP
$\mathcal{R}: \mathbb{R}^{h} \to \mathbb{R}^{K}$ produces
routing logits over $K{=}4$ latent experts. We set $K{=}4$ to
roughly match the recurring content types found in visual
documents -- running text, tables, charts, and
figures/formulas -- giving the router enough capacity to send
different regions to different specialists while keeping the
expert bank small enough to receive sufficient gradient signal
under our balanced training mixture. 
Top-$k$ sparse softmax with temperature $\tau_R$ converts these
to expert weights:
\begin{equation}
\mathbf{g}_r \;=\; \text{softmax}_k\!\left( \frac{\mathcal{R}
(\mathbf{u}_r) + \boldsymbol{\epsilon}}{\tau_R} \right),
\end{equation}
where $\tau_R = 1$ throughout training and inference,
$\boldsymbol{\epsilon} \sim \mathcal{N}(0, \sigma_R^2)$ with
$\sigma_R = 0.1$ during training and $\boldsymbol{\epsilon} = 0$
at inference, and only the top-$k$ logits are kept (the rest are
masked to $-\infty$ before the softmax).
The same $\mathbf{z}_q$ is broadcast to every region of a given
page, but each region's content $\mathbf{R}_r$ and position
$\mathbf{c}_r$ differ, so different regions receive different
routing decisions. The same region also receives different
routing for different queries, which is the property that makes
Argus query-conditioned.

\subsection{Latent Expert Bank and Shared Expert}

The latent expert bank consists of $K{=}4$ small position-wise
feed-forward modules of the form $\text{LN} \to
\text{Linear}(h, 2h) \to \text{GELU} \to \text{Linear}(2h, h)$,
each applied to $\mathbf{G}_{\text{final}}$ to produce $K$
candidate grids $\mathbf{E}_k \in \mathbb{R}^{H \times W \times h}$.
A separate \emph{shared expert} (LN, expansion factor 4, GELU,
down-projection; always active) provides a stable dense
fallback path. Region expert weights $\mathbf{g}_r$ are
broadcast from each region to its constituent patches, and the
routed specialist output is the per-patch weighted sum
\begin{equation}
\mathbf{S}_{ij} \;=\; \sum_{k=1}^{K}
g_{\,r(i,j),\,k} \,\bigl[\mathbf{E}_k(\mathbf{G}_{\text{final}})
\bigr]_{ij} .
\end{equation}
Routing is computed at the region level for stability and
cost; expert mixing is applied at the patch level so that
MaxSim retains its fine-grained evidence.

\subsection{Gated Residual Fusion}

The final grid is updated residually under learned scalar
gates:
\begin{equation}
\mathbf{F} \;=\; \mathbf{G}_{\text{final}} \;+\; \sigma(\alpha)
\,\mathbf{S}_{\text{shared}} \;+\; \sigma(\beta)\,\mathbf{S} .
\end{equation}
The two scalars are jointly learned with the rest of the model
and govern how strongly the dense fallback and the routed
specialists are mixed into the original document features.
Initializing them so that $\sigma(\alpha) = \sigma(\beta) = 0.5$
keeps Argus close to its dense ColQwen3.5 baseline at the start
of training. The fused tokens are written back into the
image-token positions of $\mathbf{H}_{\text{final}}$, and the
standard projection head produces the final document
embeddings $\mathbf{D}(q) \in \mathbb{R}^{n \times d}$, which
are L2-normalized and scored against $\mathbf{Q}$ by MaxSim.

\section{Experimental Setup}
\label{sec:experimental-setup}

\subsection{Training Objective}
\label{sec:training-objective}

The joint training phase combines two loss terms:
\begin{equation}
\mathcal{L} \;=\; \mathcal{L}_{\text{NCE}}
\;+\; \lambda_{\text{bal}} \, \mathcal{L}_{\text{bal}} .
\end{equation}

\paragraph{Retrieval (NCE).}
The retrieval term is the contrastive InfoNCE loss of
\S\ref{sec:contrastive-bg} with $\text{sim}(\cdot,\cdot)$ set to
MaxSim. For each positive pair $(q, d^+)$, the negative set
$D_N$ contains only the \emph{other documents in the same
batch}. We do not mine explicit hard negatives.
\begin{table*}[t]
\centering

\setlength{\tabcolsep}{4pt}
\caption{Retrieval accuracy (NDCG@10) on the eight public
ViDoRe V3 tasks. Baselines listed in \S\ref{sec:baselines}.}
\label{tab:v3}
\resizebox{0.9\textwidth}{!}{%
\begin{tabular}{l*{8}{c}|c}
\toprule
% & \multicolumn{8}{c|}{\textbf{Public tasks}} & \\
\textbf{Model} & CompSci & Energy & FinanceEn & FinanceFr & HR &
Industrial & Pharma & Physics & \textbf{Avg} \\
\midrule
Nemotron-colembed-8b-v2          & 79.30 & 69.82 & 67.19 & 51.54 & 66.32 & 56.03 & 67.19 & 50.84 & 63.53 \\
Tomoro-colqwen3-embed-8b         & 75.35 & 68.41 & 65.08 & 49.10 & 63.98 & 54.41 & 66.36 & 50.13 & 61.60 \\
Nemotron-colembed-4b-v2          & 78.56 & 67.48 & 65.02 & 49.01 & 62.39 & 56.10 & 64.86 & 52.78 & 62.02 \\
Ops-Colqwen3-4B                  & 77.74 & 66.49 & 65.71 & 48.81 & 61.81 & 53.99 & 66.42 & 49.14 & 61.26 \\
Tomoro-colqwen3-embed-4b         & 75.44 & 66.43 & 63.84 & 46.83 & 60.09 & 53.58 & 65.74 & 49.32 & 60.16 \\
Llama-Nemotron-colembed-3b-v2    & 77.09 & 64.88 & 64.23 & 44.41 & 62.28 & 51.71 & 66.04 & 46.93 & 59.70 \\
Jina-embeddings-v4               & 71.81 & 63.50 & 59.30 & 46.10 & 59.53 & 50.38 & 63.09 & 46.63 & 57.54 \\
ColNomic-embed-multimodal-7b     & 76.20 & 63.58 & 56.57 & 45.46 & 58.67 & 50.13 & 62.26 & 48.25 & 57.64 \\
Llama-Nemoretriever-colembed-3b-v1 & 75.16 & 62.07 & 60.88 & 43.77 & 58.69 & 47.09 & 63.74 & 45.13 & 57.07 \\
\midrule
\ours{}\textit{Argus-2B} & 75.83 & 65.82 & 64.27 & 47.62 & 62.33 & 50.76 & 64.86 & 49.27 & 60.09 \\
\ours{}\textit{Argus-4B} & 76.97 & 68.58 & 63.54 & 51.43 & 65.64 & 54.47 & 66.08 & 50.06 & 62.09 \\
\ours{}\textit{Argus-9B} & 78.03 & 69.59 & 64.74 & 50.90 & 65.46 & 53.45 & 67.29 & 50.51 & 62.50 \\
\bottomrule
\end{tabular}
}
\end{table*}
\paragraph{Load balance.}
To keep the latent expert bank well used, we adopt the
load-balance loss of \citet{switch}. Let $m_k$ be the average
softmax routing probability of expert $k$ over the valid
pooled regions in a batch and $f_k$ be the fraction of those
regions whose top-$1$ routing assignment is expert $k$. The
balance loss is
\begin{equation}
\mathcal{L}_{\text{bal}} \;=\; K \sum_{k=1}^{K} m_k \, f_k .
\end{equation}
This term is minimized when both the soft and hard routing
distributions are uniform over the $K{=}4$ latent experts,
which discourages collapse to a single specialist while
remaining compatible with sparse top-$k$ selection.

\paragraph{Warmup-only routing supervision.}
The router warmup phase
(\S\ref{sec:training-setup}, ``Router warmup'') adds a small
cross-entropy term $\mathcal{L}_{\text{rt}}$ on the routing
logits, weighted by $\lambda_{\text{rt}}$, to encourage every
latent expert to be exercised before retrieval gradients are
introduced. The retrieval term $\mathcal{L}_{\text{NCE}}$ is
disabled during warmup, and $\mathcal{L}_{\text{rt}}$ is set
to zero again in the joint phase.

\subsection{Training Data and Hyperparameters}
\label{sec:training-setup}

All three Argus sizes initialize from
\texttt{Qwen3.5-VL-\{2B,4B,9B\}-Instruct} and are trained with
PEFT/LoRA \citep{lora} of rank 32 on $4\times$ H100 80\,GB. The
training mixture contains 593,677 pairs (9.3\% of a
6,388,525-example source pool) drawn from ViDoRe ColPali, VDR
multilingual, VisRAG, TabFQuAD, TatDQA, and a capped
Docmatix-IR subset. We use $K{=}4$ latent experts with top-$2$
routing, region size $r{=}4$, and a routing tap at backbone
layer $L{-}5$. Source-level data counts and full optimizer
hyperparameters are in
Appendix~\ref{sec:appendix-training-data} and
Appendix~\ref{sec:appendix-hparams}.

\paragraph{Router warmup.}
A routed MoE module is prone to collapse if it sees retrieval
gradients before learning to use all experts, so we first run a
short warmup phase. The MoE module is attached to the backbone
with the fusion gates frozen near zero, PEFT weights frozen, and
the router and experts trained only against the routing and
balance auxiliary losses. This phase has no retrieval signal:
its only job is to make the router use all $K$ experts before
the retrieval objective is added.

\subsection{Baselines}
\label{sec:baselines}

We compare Argus against the strongest open late-interaction
VDR systems available on the public MTEB leaderboards: ColPali
\citep{faysse2024colpali}, ColSmol-500M \citep{colsmolcard},
ColQwen2 \citep{colqwen2card}, ColQwen2.5
\citep{colqwen25card}, ColNomic-3b/7b
\citep{colnomic3bcard,colnomic7bcard}, Sauerkraut-4b/8b
\citep{sauerkraut4bcard,sauerkraut8bcard}, Nemotron-colembed
\citep{nemotron3bcard,nemotron4bcard,nemotron8bcard},
Ops-Colqwen3-4B \citep{opscard}, Tomoro-colqwen3-embed-4b/8b
\citep{tomoro4bcard,tomoro8bcard}, ColNomic-embed-multimodal-7b
\citep{colnomic7bcard}, and Jina-embeddings-v4. For
MIRACL-Vision we additionally compare against the baselines
published with that benchmark
\citep{osmulsk2025miraclvision}. 
\begin{table*}[t]
\centering
\setlength{\tabcolsep}{4pt}
\caption{MIRACL-Vision results (NDCG@10) on multilingual
visual document retrieval. 
}
\label{tab:miracl_vision}
\resizebox{0.9\textwidth}{!}{%
\begin{tabular}{lcccccccc>{\columncolor{oursrow}}c}
\toprule
\textbf{Language} & \makecell{dse-qwen2\\-2b-mrl-v1} & \makecell{gme-Qwen2\\-VL-2B} & \makecell{vdr-2b\\-multi-v1} & \makecell{colqwen2\\-v1.0} & \makecell{llama-nemo\\-3b-v1} & \makecell{nemo-cb\\-3b-v2} & \makecell{nemo-vl\\-4b-v2} & \makecell{nemo-vl\\-8b-v2} & \textbf{Argus-9B} \\
\midrule
German     & 0.6267 & 0.6345 & 0.6205 & 0.5995 & 0.6831 & 0.6924 & 0.7100 & 0.7233 & \textbf{0.7236} \\
English    & 0.6605 & 0.6784 & 0.6784 & 0.6417 & 0.7363 & 0.7397 & 0.7246 & \textbf{0.7480} & 0.7149 \\
Spanish    & 0.5927 & 0.6277 & 0.6274 & 0.6224 & 0.7109 & 0.7236 & 0.7033 & 0.7089 & \textbf{0.7307} \\
Finnish    & 0.4162 & 0.6863 & 0.5283 & 0.6604 & 0.8513 & 0.8541 & 0.8398 & \textbf{0.8726} & 0.8519 \\
French     & 0.7160 & 0.6851 & 0.7194 & 0.6876 & 0.7988 & 0.7943 & 0.7943 & 0.8171 & \textbf{0.8301} \\
Indonesian & 0.4866 & 0.5416 & 0.5254 & 0.5320 & 0.6428 & 0.6550 & 0.6480 & \textbf{0.6680} & 0.6125 \\
Japanese   & 0.6232 & 0.7305 & 0.6553 & 0.6970 & 0.7260 & 0.7493 & 0.8326 & \textbf{0.8690} & 0.7750 \\
Russian    & 0.6505 & 0.7202 & 0.6995 & 0.6811 & 0.7670 & 0.7920 & 0.7879 & \textbf{0.8399} & 0.8184 \\
Yoruba     & 0.4178 & 0.4884 & 0.4577 & 0.5120 & 0.5888 & 0.5943 & 0.4469 & 0.5252 & \textbf{0.8099} \\
Chinese    & 0.5962 & 0.6314 & 0.5963 & 0.4926 & 0.4355 & 0.4878 & 0.6697 & \textbf{0.7204} & 0.6851 \\
\midrule
\textbf{Average} & 0.5786 & 0.6424 & 0.6108 & 0.6126 & 0.6941 & 0.7083 & 0.7157 & 0.7492 & \textbf{0.7552} \\
\bottomrule
\end{tabular}
}
\end{table*}

\section{Main Results}
\label{sec:main-results}

We evaluate Argus on three retrieval benchmarks and one
agentic-retrieval setting. The standalone leaderboards are the
ViDoRe V1+V2 mix (\S\ref{sec:v12}), the more recent V3 mix
(\S\ref{sec:v3}), and the multilingual MIRACL-Vision benchmark
(\S\ref{sec:miracl}). On V3 we additionally compare against the
agentic version of the same retriever (\S\ref{sec:agentic-v3}).

\subsection{ViDoRe V1+V2 Leaderboard}
\label{sec:v12}

Following the convention used by the official MTEB ViDoRe
V1+V2 leaderboard \citep{vidore-leaderboard} and prior work
\citep{moreira2026nemotron}, we report \textbf{NDCG@5} on V1
(10 in-domain tasks) and V2 (4 out-of-domain tasks) in
Table~\ref{tab:v12}. \textit{Argus-9B} attains the highest overall average (86.0),
beating \textit{nemotron-colembed-vl-8b-v2} (84.7) by
\textbf{+1.3} and \textit{Ops-Colqwen3-4B} (84.6) by +1.4. The
gap is largest on the harder out-of-domain V2 split (+4.3 over
Nemotron-8b-v2), and \emph{Argus-9B is the best system on each
of the four V2 tasks}: BiomedicalLectures, ESGReports,
ESGReports-HighLevel, and EconomicsReports. This supports the
hypothesis that adapting the document representation to the
query helps most when documents are out-of-distribution.
\textit{Argus-4B} (84.2) is competitive with the 4B SOTA
\textit{Ops-Colqwen3-4B} (84.6) and outperforms
\textit{Nemotron-4b-v2} (83.7) and \textit{Sauerkraut-4b}
(82.3). \textit{Argus-2B} (82.9) is competitive with several
much larger 7B and 8B baselines, including \textit{ColNomic-7b}
(81.5), and approaches \textit{Sauerkraut-8b} (83.0) at less
than a third of the parameter count. Argus reaches these scores with a 1024-dimensional retrieval
head and the balanced training mixture of
Appendix~\ref{sec:appendix-training-data}, narrower and smaller
than the head dimensions and training pools used by the
strongest baselines (Appendix~\ref{sec:appendix-baseline-context}),
and without seed averaging or post-training checkpoint merging.

\subsection{ViDoRe V3 Leaderboard}
\label{sec:v3}

Table~\ref{tab:v3} reports the ViDoRe V3 leaderboard restricted
to the eight public domains.
The Argus models at 2B, 4B, and 9B scale are
listed alongside the strongest published systems. Argus-9B
(62.50) is competitive with the best 8B baseline,
Nemotron-colembed-8b-v2 (63.53), at a 1024-dimensional retrieval
head versus the 4096-dimensional head used by the latter.

\subsection{Agentic ViDoRe V3 Retrieval}
\label{sec:agentic-v3}

We additionally evaluate Argus as the retriever inside an
agentic search pipeline. The agent uses a vLLM-served
Qwen3.6-27B model and exposes three tools (think, retrieve,
final-results) that let it rewrite or decompose the query,
call the late-interaction retriever multiple times, and return
a final ranked list of document ids. The reported metric is
NDCG@10 over the final ranked list; no generated answer is
scored. We set the target list size to 10 and the retriever
candidate budget to 100. Full prompts, model serving details,
and fallback logic are in
Appendix~\ref{sec:appendix-agentic}. Figure~\ref{fig:agentic-before-after} compares the same
\textit{Argus-9B} retriever before and after wrapping it
in the Qwen3.6 agent on the public ViDoRe V3 tasks
available in our run. The agent improves the macro average from
60.28 to 64.80 NDCG@10, with gains on six of the tasks.
Argus is therefore not only a strong standalone retriever, but
also a useful search primitive for iterative retrieval agents
that can refine or decompose the query.

\subsection{Multilingual MIRACL-Vision}
\label{sec:miracl}

Table~\ref{tab:miracl_vision} reports NDCG@10 on
MIRACL-Vision, alongside the
baselines published with that benchmark.  Argus reaches the
highest macro average across the ten languages (0.7552 versus
0.7492 for Nemotron-colembed-8b-v2) and is the best system on
five of the ten languages, including the long-tail Yoruba split
(0.8099 versus 0.5252 for Nemotron-8b-v2). This complements the
V1+V2 picture by showing that the gains from query-conditioned
routing carry over to a different multilingual evaluation
distribution.

\section{Analysis}
\label{sec:analysis}

The previous section established the headline retrieval
numbers. This section unpacks them along three axes: how the
gain distributes across the 2B/4B/9B family
(\S\ref{sec:scaling}), which design choices of the
query-conditioned MoE module carry the gain
(\S\ref{sec:ablations}), and what the learned router actually
does (\S\ref{sec:routing-analysis}). %All analyses use the
%released checkpoints; no additional training is performed.

\subsection{Scaling Across the Argus Family}
\label{sec:scaling}

Argus improves from 82.9 to 86.0 average NDCG@5 as the backbone
grows from 2B to 9B (Table~\ref{tab:v12}). The main change is on
ViDoRe V2: the 4B to 9B step adds +5.1 points on V2, while V1
changes only modestly. Larger backbones therefore help the
query-conditioned router most under domain shift, and the V1
ceiling is approximately reached already at the 4B scale. A
visualization of this trend is given in
Appendix~\ref{sec:appendix-scaling-vis}
(Figure~\ref{fig:scaling}).

\subsection{Which Components Carry the Gain?}
\label{sec:ablations}

Table~\ref{tab:ablation} ablates the three signals that the
query-aware router combines: the query context
$\mathbf{z}_q$, the region position $\mathbf{c}_r$, and the
always-active shared expert. All ablations are run on the
Argus-4B configuration and reported as NDCG@5 on the V1 and V2
splits. Each ablated variant is retrained from the same
Qwen3.5-VL-4B initialization, with the same training data
mixture (\S\ref{sec:appendix-training-data}), the same
hyperparameters (\S\ref{sec:appendix-hparams}), and the same
two-phase router-warmup-then-joint schedule as the released
Argus-4B model; only the listed component is removed before
training begins. Removing any of the three components drops V1+V2 average by
roughly 3 to 4 points, with the largest losses on the
out-of-domain V2 split, confirming that all three pieces are
load-bearing. The hypothesis behind each ablation row is given
in Appendix~\ref{sec:appendix-ablation-hypotheses}.
\begin{figure}[t]
\centering
\includegraphics[width=0.9\linewidth]{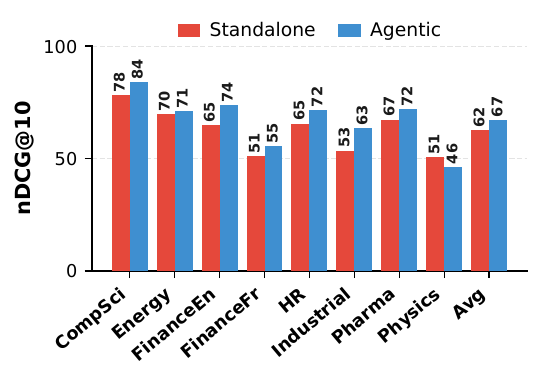}
\caption{Standalone versus agentic Argus retrieval on the
 public ViDoRe V3 tasks. }
\label{fig:agentic-before-after}
\end{figure}
\begin{table}[t]
\centering
\small
\setlength{\tabcolsep}{4pt}
\caption{Argus-4B component ablation (NDCG@5). Each row
retrains the full model with the listed component removed.}
\label{tab:ablation}
\begin{tabular}{lcc}
\toprule
\textbf{Configuration} & \textbf{V1} & \textbf{V2} \\
\midrule
\ours{}\textit{Argus-4B} (full) & 92.3 & 64.1 \\
\midrule
\;\;$-$ query conditioning ($\mathbf{z}_q$) & 88.9 & 59.5 \\
\;\;$-$ region position ($\mathbf{c}_r$)    & 88.8 & 59.4 \\
\;\;$-$ shared expert ($\sigma(\alpha) = 0$) & 89.7 & 59.1 \\
\bottomrule
\end{tabular}
\end{table}

\subsection{What Does the Router Learn?}
\label{sec:routing-analysis}

\begin{figure}[t]
  \centering
  \includegraphics[width=0.9\linewidth]{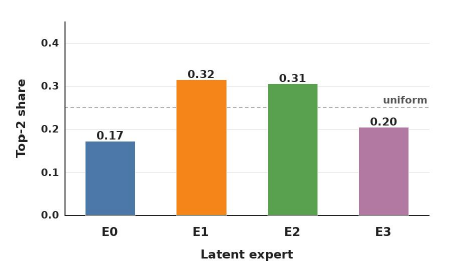}
  \caption{\textbf{Expert utilization on ViDoRe V2.} Average
  share of regions that select each latent expert in their
  top-$2$ set. }
  \label{fig:expert-load}
\end{figure}

The router is explicitly conditioned on the query context
$\mathbf{z}_q$, so the same page can receive different sparse
routing patterns for different queries. On a mixed-content
ViDoRe V2 Economics page, the released Argus-4B reroutes 24 of
63 pooled regions to a different top-1 expert when the query
changes; a side-by-side visualization of this example is given
in Appendix~\ref{sec:appendix-routing-vis}
(Figure~\ref{fig:routing-heatmap}). We treat this as qualitative
evidence only; the controlled test of query conditioning is the
ablation in Table~\ref{tab:ablation}, where $\mathbf{z}_q$ is
removed while the rest of the model is kept fixed. Aggregate top-$2$ utilization across the same ViDoRe V2
Economics sample is shown in Figure~\ref{fig:expert-load}; all
four latent experts remain active at the corpus level. The dashed line marks the uniform $1/K$ reference. The measured shares are $(0.17,0.32,0.31,0.20)$, with entropy 1.36 and 3.88 effective experts. This aggregate view complements the single-query sharp routing pattern in Figure~\ref{fig:routing-heatmap}. We additionally sweep the expert-bank size and the router tap
depth in Appendix~\ref{sec:appendix-num-experts} and
Appendix~\ref{sec:appendix-router-depth}. Both ablations show
an inverted-U with a peak at the released configuration
($K{=}4$ experts, router tap $L{-}5$), and both confirm that
V2 is the more sensitive split.

\subsection{Index Storage}
\label{sec:deployment}
\label{sec:storage}

Late-interaction retrievers must store \emph{all} per-token
document embeddings. Per-page index size therefore scales
linearly with both the number of visual tokens and the
embedding dimension. We compute storage as
\[
\mathrm{MB/page} =
\frac{\text{tokens/page}\times d \times 2}{10^6},
\]
assuming bf16 token embeddings. Argus deliberately keeps the
embedding dimension fixed at $d{=}1024$ for all released
sizes. This is narrower than the 2560-dimensional heads used by
Ops-Colqwen3-4B and Nemotron-4B-v2, and the 4096-dimensional
head used by Nemotron-8B-v2. Since token budgets differ across
systems, Table~\ref{tab:storage} reports both axes explicitly
rather than comparing dimensions alone. Under documented
per-page token budgets, Argus-9B uses $4.5\times$ less storage
than Nemotron-colembed-8b-v2 ($2.8\times$ less than
Nemotron-colembed-4b-v2) while reaching the highest V1+V2
average; the storage--accuracy frontier is visualized in
Appendix~\ref{sec:appendix-inference}
(Figure~\ref{fig:storage-pareto}).

\begin{table}[t]
\centering
\small
\setlength{\tabcolsep}{4pt}
\caption{Per-page index storage in bf16, using each model's
documented per-page token budget.}
\label{tab:storage}
\resizebox{0.4\textwidth}{!}{%
\begin{tabular}{lccr}
\toprule
\textbf{Model} & \textbf{Tok/page} & \textbf{Dim} & \textbf{MB/page} \\
\midrule
Tomoro-colqwen3-4b      & 1280 & 320  & 0.8 \\
\ours{}\textit{Argus-2B} & 2048 & 1024 & 4.2 \\
\ours{}\textit{Argus-4B} & 2048 & 1024 & 4.2 \\
Ops-Colqwen3-4B         & 1280 & 2560 & 6.6 \\
Nemotron-colembed-4b-v2 & 2304 & 2560 & 11.8 \\
\midrule
\ours{}\textit{Argus-9B} & 2048 & 1024 & 4.2 \\
Nemotron-colembed-8b-v2 & 2304 & 4096 & 18.9 \\
\bottomrule
\end{tabular}}
\end{table}

\section{Conclusion}

Argus introduces a query-conditioned region-aware
MoE module into the late-interaction VLM
retrieval pipeline, allowing the same document page to be
processed differently for different queries while preserving
all of the retrieval-side properties of ColPali-style
multi-vector indexing. Across the Argus family (2B/4B/9B), the
9B model achieves a new state of the art on ViDoRe V1+V2 among
open late-interaction models, with the largest gains on
out-of-domain V2 tasks and a favorable storage-accuracy trade
off.

\section*{Limitations}

Argus inherits the storage and per-query latency overheads of
late interaction. The query-conditioned router additionally
prevents the standard ColPali optimization of indexing
$\mathbf{D}$ once per corpus, so deployment uses a two-stage
retrieve-then-rerank pipeline. Our V3 evaluation also covers
only the eight public domains; the private Nuclear and Telecom
splits are held out by the MTEB maintainers and are not
reported. The agentic V3 results in
Figure~\ref{fig:agentic-before-after} are limited to the
public tasks our run reached, and use a single Qwen3.6 27B
agent at fixed search budgets; results may differ with other
backbones or budgets. Finally, the latent experts are not
constrained to interpretable categories such as ``table'' or
``chart''. Specialization is implicit and discovered during
training, which limits the interpretability of individual
routing decisions.

\bibliography{custom}

\appendix

% \clearpage
\appendixpage            % Optional: Creates a title page for the appendix
\addappheadtotoc         % Optional: Adds "Appendix" to the TOC
\numberwithin{figure}{section}
\numberwithin{table}{section}

This appendix provides supplementary material for Argus. It is
organized as follows:

\begin{itemize}
\item Appendix~\ref{sec:appendix-training-data} reports the
source-level training data mixture used by all released Argus
checkpoints.
\item Appendix~\ref{sec:appendix-hparams} lists the optimizer
and routing hyperparameters for the warmup and joint training
phases.
\item Appendix~\ref{sec:appendix-baseline-context} summarizes
the head dimensions, training pool sizes, and post-training
tricks used by the strongest open baselines on the V1+V2
leaderboard.
\item Appendix~\ref{sec:appendix-inference} details the
offline-cache inference pipeline and reports per-page encoding
and per-query latency measurements.
\item Appendix~\ref{sec:appendix-singlevec} compares Argus-9B
under multi-vector late interaction against average-pooled
single-vector scoring on V1+V2.
\item Appendix~\ref{sec:appendix-ablation-hypotheses} states
the hypothesis behind each row of the component ablation in
Table~\ref{tab:ablation}.
\item Appendix~\ref{sec:appendix-routing-vis} shows a same-page
two-query routing visualization on a ViDoRe V2 Economics page.
\item Appendix~\ref{sec:appendix-scaling-vis} visualizes the
V1 / V2 scaling trend across the Argus 2B/4B/9B family.
\item Appendix~\ref{sec:appendix-agentic} gives the full agent
prompts and serving protocol used in the agentic ViDoRe V3
evaluation.
\item Appendix~\ref{sec:appendix-num-experts} sweeps the
latent expert-bank size $K$ on Argus-4B.
\item Appendix~\ref{sec:appendix-router-depth} sweeps the
router tap depth on Argus-4B.

\end{itemize}

\paragraph{Key Tables}
\begin{itemize}
\item Table~\ref{tab:training-data}: training mixture after
source balancing (593{,}677 pairs, 9.3\% of the source pool).
\item Table~\ref{tab:hparams}: training hyperparameters for the
warmup and joint phases.
\item Table~\ref{tab:inference-time}: per-page offline encoding
cost and per-query online retrieval latency on ViDoRe V2 ESG
Reports.
\item Table~\ref{tab:inference-family}: family-wide encoding
and latency measurements on V2 Economics Reports.
\item Table~\ref{tab:singlevec}: Argus-9B under multi-vector
late interaction vs.\ average-pooled single-vector scoring.
\item Table~\ref{tab:ablation-experts}: expert-count
ablation on Argus-4B.
\item Table~\ref{tab:ablation-router-depth}: router tap-depth
ablation on Argus-4B.
\end{itemize}

\begin{figure}[h]
  \centering
  \includegraphics[width=\linewidth]{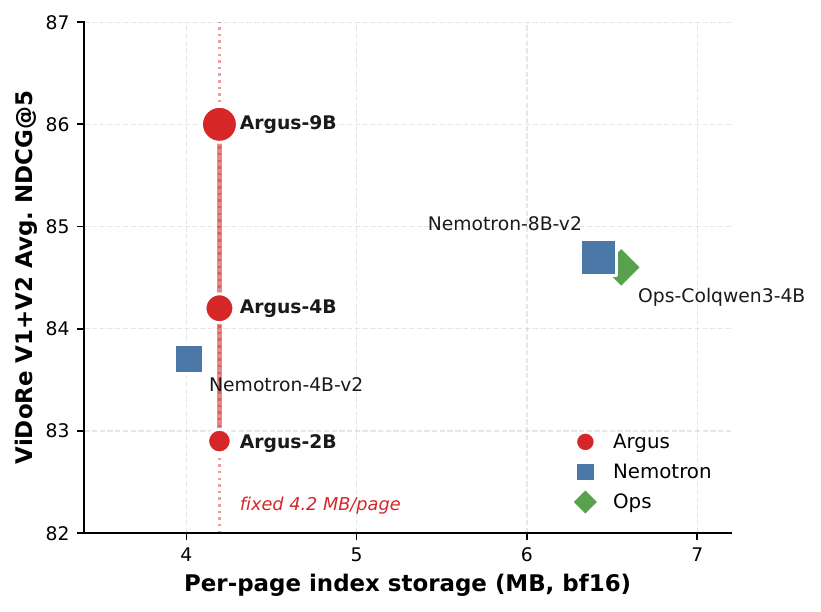}
\caption{\textbf{Storage and accuracy trade-off.} Per-page
  bf16 index size versus ViDoRe V1+V2 average NDCG@5 for models
  where both storage settings and V1+V2 results are available.
  Argus-9B improves accuracy while keeping the same 4.2 MB/page
  index footprint as the smaller Argus models.}
  \label{fig:storage-pareto}
\end{figure}

\paragraph{Key Figures}
\begin{itemize}
\item Figure~\ref{fig:inference}: the Argus inference pipeline
showing offline document caching and the per-query online path.
\item Figure~\ref{fig:storage-pareto}: storage--accuracy
trade-off across systems with both storage and V1+V2 results
available.
\item Figure~\ref{fig:routing-heatmap}: same-page,
different-query routing visualization on a ViDoRe V2 Economics
page.
\item Figure~\ref{fig:scaling}: V1/V2 scaling of the Argus
family.
\item Figure~\ref{fig:prompt-agentic-main}: main prompt used
by the Qwen3.6-27B agent.
\item Figure~\ref{fig:prompt-agentic-selection}: selection-
agent prompt used as an auxiliary reranking stage.
\end{itemize}

\section{Training Data Mixture}
\label{sec:appendix-training-data}

Table~\ref{tab:training-data} gives the source-level training
mixture used by the released Argus checkpoints. The main paper
reports only the aggregate size; the table is included here for
reproducibility.

\begin{table}[h]
\centering
\footnotesize
\setlength{\tabcolsep}{3pt}
\caption{Training mixture after source balancing. The final
training set uses 593,677 examples, or 9.3\% of the full
available pool.}
\label{tab:training-data}
\begin{tabular}{lrr}
\toprule
\textbf{Source} & \textbf{Full} & \textbf{Used} \\
\midrule
visrag\_synthetic & 239,206 & 60,000 \\
visrag\_in\_domain & 122,752 & 60,000 \\
tatdqa\_train & 13,251 & 13,251 \\
tabfquad\_train & 1,552 & 1,552 \\
vdr\_multilingual\_en & 53,512 & 53,512 \\
vdr\_multilingual\_de & 58,217 & 58,217 \\
vdr\_multilingual\_fr & 55,270 & 55,270 \\
vdr\_multilingual\_es & 58,738 & 58,738 \\
vdr\_multilingual\_it & 54,942 & 54,942 \\
colpali\_arxiv\_qa & 9,949 & 9,949 \\
colpali\_docvqa & 39,302 & 39,302 \\
colpali\_infographic\_vqa & 10,038 & 10,038 \\
colpali\_pdf & 45,718 & 45,718 \\
colpali\_tatdqa & 13,188 & 13,188 \\
docmatix\_resolved & 5,612,890 & 60,000 \\
\midrule
\textbf{Total} & \textbf{6,388,525} & \textbf{593,677} \\
\bottomrule
\end{tabular}
\end{table}

\section{Training Hyperparameters}
\label{sec:appendix-hparams}

Table~\ref{tab:hparams} summarizes the training hyperparameters
used for the released Argus checkpoints. The same values are
used at every scale (2B, 4B, 9B); only the per-device batch
size and the final data mixture differ. All weights are saved in
fp32 with merged PEFT after training.

\begin{table}[h]
\centering
\small
\setlength{\tabcolsep}{4pt}
\caption{Training hyperparameters for the released Argus
checkpoints. The warmup phase trains only the router and
experts against the auxiliary losses; the joint phase
unfreezes PEFT and the fusion gates and adds the retrieval
objective.}
\label{tab:hparams}
\begin{tabular}{lcc}
\toprule
\textbf{Hyperparameter} & \textbf{Router warmup} & \textbf{Joint} \\
\midrule
PEFT/LoRA rank          & frozen  & 32  \\
Train MoE gates         & frozen  & yes \\
$K$ (latent experts)    & 4       & 4   \\
Top-$k$ routing         & 2       & 2   \\
Region size $r$         & 4       & 4   \\
Router layer            & $L{-}5$ & $L{-}5$ \\
Router temperature      & 1.0     & 1.0 \\
Router noise std        & 0.1     & 0.1 \\
$\lambda_{\text{rt}}$   & 0.1     & 0.0 \\
$\lambda_{\text{bal}}$  & 0.5     & 0.5 \\
Negatives per sample    & 2       & 2 \\
Learning rate           & $1\!\times\!10^{-5}$ & $5\!\times\!10^{-6}$ \\
Effective batch size    & 64      & 64 \\
Precision               & bf16/fp32 & bf16/fp32 \\
\bottomrule
\end{tabular}
\end{table}

\section{Baseline Configuration Context}
\label{sec:appendix-baseline-context}

For context on the V1+V2 leaderboard comparison in
\S\ref{sec:v12}, the strongest open baselines use the following
design choices. The Nemotron ColEmbed V2 cards report
4096-dimensional outputs for the 8B model and 2560-dimensional
outputs for the 4B model, with model merging and an
approximately 500k-image training mixture
\citep{nemotron8bcard,nemotron4bcard}. Ops-Colqwen3-4B reports
its strongest V1+V2 setting at 2560 dimensions
\citep{opscard}. The webAI 9B card reports a 2560-dimensional
projection head and roughly 2M question-image training pairs
\citep{webaicard}, while the athrael-soju ColQwen3.5 card
reports a 320-dimensional trained head, roughly 776k pairs,
three training seeds, seed averaging, and model soup
\citep{athraelcard}. Argus uses a 1024-dimensional head at every size and the balanced training mixture
summarized in Appendix~\ref{sec:appendix-training-data}, without
seed averaging or post-training checkpoint merging.

\section{Query-Time Inference Cost}
\label{sec:appendix-inference}

Although the Argus document representation is
query-conditioned, the per-query cost can be made comparable to
that of a query-independent late-interaction retriever by
caching all query-independent visual features once per corpus
(\S\ref{sec:deployment}). To verify that the added router and
fusion path does not impose a large query-time penalty, we
measure end-to-end retrieval latency on a fixed set of 50
ViDoRe V2 queries.

\begin{figure*}[h]
\centering
\includegraphics[width=\linewidth]{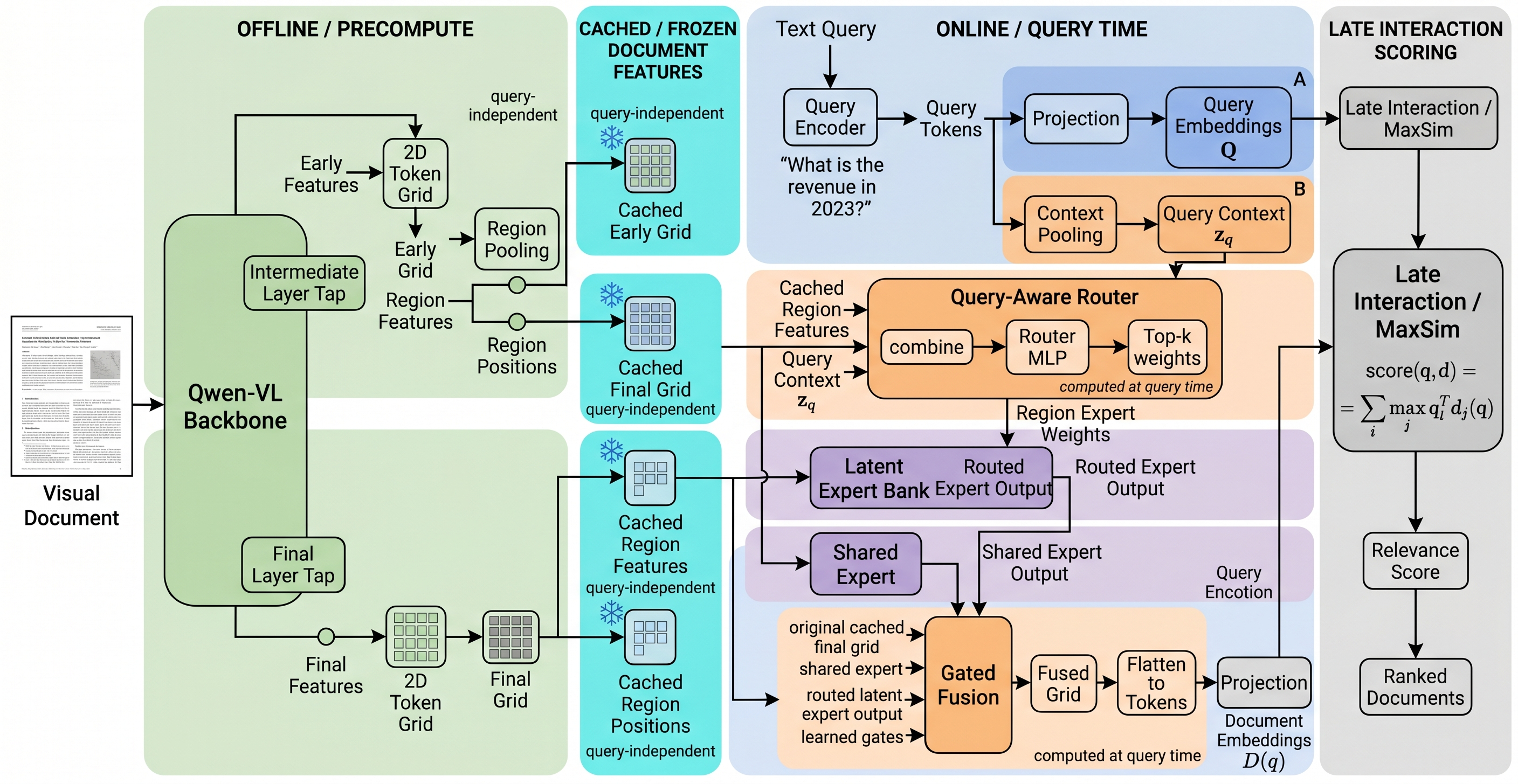}
\caption{\textbf{Argus inference pipeline.} The offline
precompute pass caches all query-independent document features
(early grid, final grid, region features, region positions, and
optionally the latent expert outputs). The document image encoder
is not run at query time. The online path encodes the text query,
routes cached regions with $\mathbf{z}_q$, fuses cached visual
features, projects the resulting tokens, and applies MaxSim to
candidate pages, producing the query-conditioned document
embeddings $\mathbf{D}(q)$.}
\label{fig:inference}
\end{figure*}

Figure~\ref{fig:inference} separates the offline and online
parts of the computation. During offline indexing, each page is
passed through the Qwen-VL document image encoder once. This
produces the intermediate grid used by the router, the final
grid used by the expert bank, pooled region descriptors, and
region coordinates. None of these cached tensors depends on the
query.

At query time, Argus does not concatenate the query with the
image and does not rerun the image encoder. The text query is
encoded once to obtain query tokens $\mathbf{Q}$ and the routing
context $\mathbf{z}_q$. A static first-stage retriever provides a
small candidate set. For each candidate page, the router combines
$\mathbf{z}_q$ with cached region descriptors and coordinates,
produces top-$k$ expert weights, and applies gated fusion to the
cached final grid. The projection head then produces
$\mathbf{D}(q)$ for MaxSim scoring. Thus, query conditioning is
implemented after visual feature extraction, not by repeating
the full document encoder per query.

There are two practical caching modes. In the default mode, the
latent experts are applied online to the cached final grid for
the candidate pages only. In the larger-storage mode, the
outputs of all latent experts are also cached offline, so online
document work reduces to router weights, weighted summation,
gated residual fusion, projection, and MaxSim. In both modes,
the cost scales with the candidate set size rather than with the
full corpus.

\paragraph{Protocol.}
We use the ViDoRe V2 ESG Reports task as a stable,
leaderboard-equivalent benchmark: the same 452-page corpus and
232 queries that produce the V2 ESG score in
Table~\ref{tab:v12}. For each model we take the first 100
corpus pages and the first 55 queries (5 discarded as warmup,
50 timed). The document encoder is run once per page and the
resulting multi-vector tensor is cached. The online path
encodes the query text and runs MaxSim against the cached
document embeddings. We synchronize the GPU before and after
each timed region.
All measurements use a single NVIDIA H100 80\,GB GPU, bf16
inference, batch size 1 (worst case for latency),and the same
Python environment (PyTorch 2.11, transformers 5.7, MTEB 2.12,
flash-attn 2.6) for both models.

\begin{table}[h]
\centering
\small
\setlength{\tabcolsep}{3pt}
\caption{Per-page offline encoding cost and per-query online
retrieval latency on the ViDoRe V2 ESG Reports task (first
100 pages, 50 timed queries), measured on one H100 80\,GB at
batch 1.}
\label{tab:inference-time}
\begin{tabular}{lcc}
\toprule
& Argus-9B & \makecell{Nemotron-\\colembed-8b-v2} \\
\midrule
Doc encode (ms / page, offline)     & 373.9  & 5090.1 \\
\midrule
Query online, mean (ms)             & 135.8 & 277.6 \\
Query online, std  (ms)             &  16.2 &  17.3 \\
Query online, p50  (ms)             & 124.7 & 268.2 \\
Query online, p95  (ms)             & 161.1 & 305.9 \\
\bottomrule
\end{tabular}
\end{table}

On this leaderboard-equivalent V2 ESG corpus, Argus-9B is
$13.6\times$ faster than Nemotron-colembed-8b-v2 at offline
document encoding (374 vs.\ 5090 ms per page) and $2.0\times$
faster per query at online retrieval (136 vs.\ 278 ms). The
adaptive-tiling regime that Nemotron uses to handle large
PDF-style pages amplifies the cost gap on visually rich
documents; Argus's fixed token budget keeps both the offline
index and the per-query MaxSim cost bounded. Together with the
$4.5\times$ storage reduction in Table~\ref{tab:storage}, this
makes the deployment trade-off favorable for Argus on the V2
distribution where its accuracy advantage is also the largest.

\paragraph{Family-wide scaling.}
For completeness, we also measure offline and online latency
across the full released Argus family on the V2 Economics
Reports task (the same 100-page / 50-timed-query protocol as
above). The query text load
differs slightly from the ESG split, so the absolute numbers in
Table~\ref{tab:inference-family} are not directly comparable to
Table~\ref{tab:inference-time}; the goal here is only to track
how the cost moves with model size on a fixed task.

\begin{table}[h]
\centering
\small
\setlength{\tabcolsep}{2pt}
\caption{Per-page encoding cost and per-query latency for the
Argus family on the V2 Economics Reports task (first 100
pages, 50 timed queries), same single-H100 / batch-1 protocol
as Table~\ref{tab:inference-time}.}
\label{tab:inference-family}
\begin{tabular}{lccc}
\toprule
\textbf{Model} & \textbf{Doc/page (ms)} & \textbf{Query mean (ms)}
& \textbf{Query p50 (ms)} \\
\midrule
\ours{}\textit{Argus-2B} & 184.1 & 102.6 & 101.1 \\
\ours{}\textit{Argus-4B} & 226.1 & 131.5 & 130.3 \\
\ours{}\textit{Argus-9B} & 231.0 & 132.8 & 131.5 \\
\bottomrule
\end{tabular}
\end{table}

\section{Single-Vector vs.\ Late-Interaction Scoring}
\label{sec:appendix-singlevec}

To help understand the retrieval-accuracy difference obtained
with a single pooled vector and with late-interaction scoring,
we evaluate the same released Argus-9B checkpoint under both
regimes, using the same processor and the same MTEB ViDoRe V1
and V2 evaluation protocol used for Table~\ref{tab:v12}. In the
single-vector regime, the per-token document and query
embeddings are average-pooled along the token axis and
L2-normalized, and relevance is computed by cosine similarity
between the resulting single vectors; in the late-interaction
regime, MaxSim is applied to the same multi-vector outputs as in
the main paper. No retraining is performed.

\begin{table}[h]
\centering
\small
\setlength{\tabcolsep}{4pt}
\caption{Argus-9B retrieval accuracy (NDCG@5) under
multi-vector late interaction vs.\ average-pooled
single-vector scoring, on ViDoRe V1+V2 with the same
checkpoint and evaluation protocol.}
\label{tab:singlevec}
\begin{tabular}{lccc}
\toprule
\textbf{Argus-9B} & \textbf{V1} & \textbf{V2} & \textbf{V1+V2} \\
\midrule
Single vector     & 82.7 & 59.2 & 76.0 \\
Late interaction  & 92.7 & 69.2 & 86.0 \\
\midrule
\% improvement    & +12.1\% & +16.9\% & +13.2\% \\
\bottomrule
\end{tabular}
\end{table}

The single-vector mode trades roughly $13\%$ relative NDCG@5
for a $\sim$$2000\times$ reduction in per-page storage and a
correspondingly cheap ANN index. The V2 gap is larger than the
V1 gap (+16.9\% vs.\ +12.1\%), which is consistent with the
observation that out-of-domain pages have evidence spread over
many small regions and benefit most from the per-token matching
that late interaction enables.

\section{Ablation Component Hypotheses}
\label{sec:appendix-ablation-hypotheses}

Each row of Table~\ref{tab:ablation} isolates one hypothesis
behind the Argus design.

\paragraph{$-$ query conditioning ($\mathbf{z}_q$).}
Drops $\mathbf{W}_{\text{query}}\mathbf{z}_q$ from the router
input, so routing depends only on region content and position.
This measures the contribution of query-conditioned routing,
the central novelty of Argus.

\paragraph{$-$ region position ($\mathbf{c}_r$).}
Drops $\mathbf{W}_{\text{coord}}\mathbf{c}_r$, so the router
sees content and query but not location, testing the value of
positional bias.

\paragraph{$-$ shared expert ($\sigma(\alpha){=}0$).}
Forces $\sigma(\alpha){=}0$, so the entire expert contribution
must flow through routed latent experts. This removes the
always-on dense fallback path that stabilizes early training.

\section{Per-Query Routing Visualization}
\label{sec:appendix-routing-vis}

\begin{figure*}[h]
  \centering
  \includegraphics[width=\linewidth]{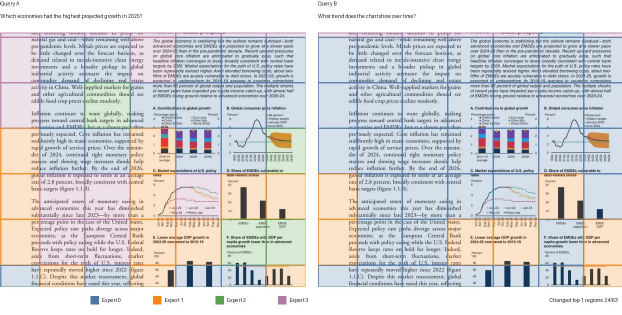}
  \caption{\textbf{Same page, different query contexts.} Top-1
  expert assignment per region of an identical ViDoRe V2
  Economics page under two queries. The routing pattern changes
  from top-1 expert counts of $(18,17,13,15)$ to
  $(39,14,4,6)$, and 24 of 63 regions change their top-1 expert.
  This single example illustrates query-sensitive routing, but
  does not assign fixed semantic labels to experts.}
  \label{fig:routing-heatmap}
\end{figure*}

Figure~\ref{fig:routing-heatmap} accompanies the qualitative
claim in \S\ref{sec:routing-analysis} that the same document
page can receive different sparse routing patterns under
different queries. We treat the example as illustrative; the
controlled test of query conditioning is the ablation in
Table~\ref{tab:ablation}.

\section{Scaling Visualization}
\label{sec:appendix-scaling-vis}

\begin{figure}[h]
  \centering
  \includegraphics[width=\linewidth]{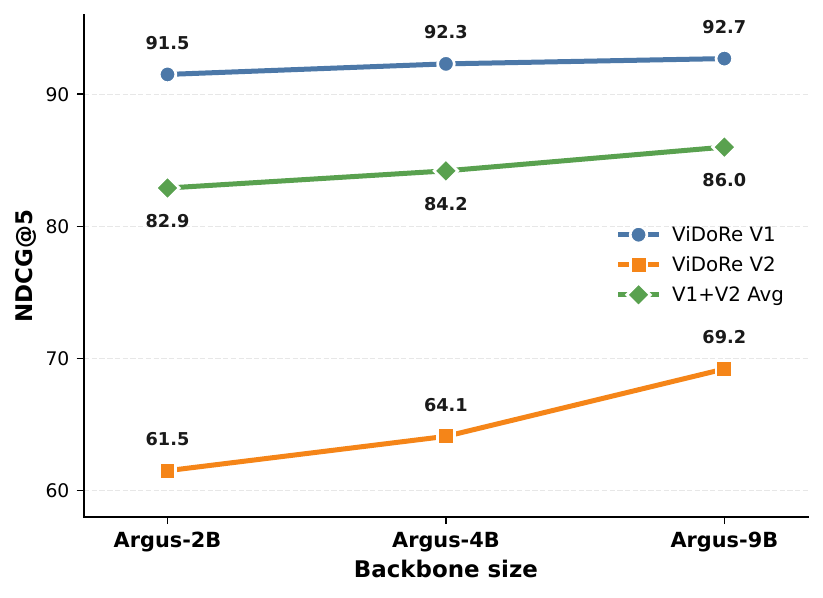}
  \caption{\textbf{Scaling of the Argus family on ViDoRe
  V1+V2.} The 4B$\rightarrow$9B step is concentrated in the
  harder out-of-domain V2 average (+5.1), while in-domain V1
  saturates around 92.7.}
  \label{fig:scaling}
\end{figure}

Figure~\ref{fig:scaling} visualizes the scaling trend
discussed in \S\ref{sec:scaling}: ViDoRe V1 saturates near 92.7
at the 4B scale, while the V2 (out-of-domain) average continues
to grow with backbone size.

\section{Agentic Retrieval Evaluation Details}
\label{sec:appendix-agentic}

The agentic results in Figure~\ref{fig:agentic-before-after}
evaluate Argus as the retrieval engine inside our agentic
search pipeline. The purpose of this experiment is to test
whether a strong late-interaction retriever remains useful when
the query is no longer a single fixed string, but is instead
reformulated and expanded by an LLM during search. The agent is
given the user query, an initial Argus ranking, and a small set
of retrieval tools. It can issue additional searches, inspect
the returned document snippets, and then commit to a final
ranked list. The evaluation uses only the final list of document
ids, so the reported score measures retrieval quality rather
than answer generation.

\paragraph{Model serving and evaluation protocol.}
We serve Qwen3.6-27B with vLLM~\cite{kwon2023efficient} on two H100 GPUs using tensor
parallelism 2, a 65{,}536 token context window, and the Qwen3
reasoning and XML tool-call parsers. Argus runs on a separate
H100 GPU and serves all retrieval calls made by the agent.
Each run uses a target list size of 10, a retriever candidate
budget of 100, concurrency 4, high reasoning effort, and the
corresponding ViDoRe V3 language filter. The metric is
NDCG@10 over the final document ranking returned by the agent.
No generated answer is evaluated.

\paragraph{Agent state and actions.}
The agent starts from the original query and the first Argus
retrieval result list. It may then run for at most 200 LLM
steps. At each step, the LLM can choose one of three actions:
retrieve, think, or final-results. The retrieve action sends a
new text query to Argus and returns document ids, retrieval
scores, and markdown snippets. This allows the agent to search
for alternative phrasing, missing entities, related
terminology, or subquestions that are implicit in the original
query. The implementation tracks which documents have already
been shown to the agent, so repeated documents are not resent
with full content. The think action records intermediate
reasoning without calling the retriever. The final-results
action ends the trajectory and must return exactly ten document
ids sorted by decreasing relevance.

\paragraph{Final ranking extraction.}
The primary output is the document list passed to
final-results. This list is used directly whenever it is
valid. If an interaction does not produce a valid final tool
call, we apply deterministic fallbacks so that every query still
has a reproducible ranking: first reciprocal-rank fusion over
all retrieval calls in the trajectory, then the auxiliary
selection output, and finally the initial Argus ranking if the
LLM call fails. The run dictionary assigns larger scores to
higher-ranked final document ids, and the ViDoRe evaluator
computes NDCG@10 from this ranked list.

\begin{figure*}[ht]
\centering
\begin{subfigure}[t]{\textwidth}
\footnotesize
\centering
\begin{tcolorbox}[width=\linewidth,
                  colback=blue!0!white, colframe=orange!75!black,
                  title=Agentic retrieval main prompt,
                  fonttitle=\bfseries]
\textbf{System Prompt:} You are a retrieval agent that finds all
documents related to a given query.

\vspace{0.5em}

\textbf{Goal.} You are given a search query and a list of
documents retrieved for that query. Your task is to write new
queries and use the search tool to find all related and somewhat
related documents to the given query, with the objective of
maximizing recall. If the user's query is a question, do not
answer the question yourself. Instead, find the related
documents for the query.

\textbf{Relevance definition.}
\begin{itemize}
\item The meaning of query, document, and relevance can be more
complex than in standard web search.
\item A relevant document is one that is useful for a user who
is searching for the given query.
\item The agent should analyze the query and available
documents, then reason about what relevance means for this
specific task.
\item If the query is itself a prompt for an LLM, relevant
documents are those that help the LLM answer or solve it.
\end{itemize}

\textbf{Workflow.}
\begin{itemize}
\item The agent is given a retrieval tool powered by a dense
embedding model.
\item The agent may call the retrieval tool multiple times.
\item The agent should search from different angles and revise
queries based on documents found in earlier steps.
\item Once confident that the relevant and somewhat relevant
documents have been found, the agent must call
\texttt{final\_results}.
\item The \texttt{final\_results} call must contain exactly ten
document ids.
\item The returned document ids must be sorted in decreasing
relevance to the query.
\end{itemize}

\textbf{Best practices.}
\begin{itemize}
\item Be thorough and find all related and somewhat related
documents.
\item Prioritize recall: if multiple documents are relevant,
find and report all of them, even if only a subset is enough to
answer the query.
\item Use the initially retrieved documents to decide what
additional queries could reveal missing relevant documents.
\end{itemize}

\textbf{Available tools.}
\begin{itemize}
\item \texttt{retrieve(query, top\_k)}: search the corpus with
the active late-interaction retriever.
\item \texttt{think(thought)}: log intermediate reasoning
without retrieving new information.
\item \texttt{final\_results(doc\_ids, message,
search\_successful)}: return the final ranked list and end the
interaction.
\end{itemize}
\end{tcolorbox}
\end{subfigure}
\caption{Main prompt used by the Qwen3.6 27B agent in the
agentic ViDoRe V3 retrieval evaluation. The prompt asks the
agent to treat retrieval as a recall-oriented search problem,
use Argus for additional query formulations when needed, and
return a ranked top-ten document list.}
\label{fig:prompt-agentic-main}
\end{figure*}

\begin{figure*}[ht]
\centering
\begin{subfigure}[t]{\textwidth}
\footnotesize
\centering
\begin{tcolorbox}[width=\linewidth,
                  colback=blue!0!white, colframe=orange!75!black,
                  title=Agentic retrieval selection prompt,
                  fonttitle=\bfseries]
\textbf{System Prompt:} You are a document re-ranker agent,
which is the final stage in an information retrieval pipeline.

\vspace{0.5em}

\textbf{Role.} You are given a search query and a list of
retrieved candidate documents that are potentially relevant to
the query. Your goal is to identify the most relevant documents
from the candidate list.

\textbf{Workflow.}
\begin{itemize}
\itemsep0pt\parsep0pt
\item Read the query carefully and understand the user's
information need.
\item Compare the query with each candidate document and judge
how useful the document is for the query.
\item Select the required number of most relevant candidate
documents.
\item Sort the selected documents by decreasing relevance.
\item Call \texttt{log\_selected\_documents} to report the final
selection and finish the task.
\end{itemize}

\textbf{Thinking tool.}
\begin{itemize}
\itemsep0pt\parsep0pt
\item Use \texttt{think} for complex analysis, ambiguous
queries, difficult relevance judgments, and ordering selected
documents.
\item The \texttt{think} tool does not retrieve new information;
it only records reasoning for transparency.
\end{itemize}
\end{tcolorbox}
\end{subfigure}
\caption{Selection-agent prompt used as an auxiliary reranking
stage. The pipeline computes top-five and top-ten selections,
but the primary evaluated output is the final-results tool
call from the main agent; selection results are used only
as a later fallback when extracting a final ranked list.}
\label{fig:prompt-agentic-selection}
\end{figure*}

\section{Number of Experts Ablation}
\label{sec:appendix-num-experts}

Argus uses $K{=}4$ latent experts with top-$2$ routing. This
section sweeps the expert-bank size while holding everything
else fixed: the Argus-4B initialization, training mixture
(\S\ref{sec:appendix-training-data}), warmup-then-joint
schedule, and router tap at $L{-}5$. Only $K$ and the top-$k$
selection are changed; each variant is retrained from scratch.
The hypothesis behind $K{=}4$ is that visual documents have
roughly four recurring content modes (running text, tables,
charts, and figures/formulas), so $K{=}4$ should give the
router enough capacity to specialize without starving any
expert of gradient signal under our $\sim$594k-pair training
mixture.

\begin{table}[h]
\centering
\small
\setlength{\tabcolsep}{4pt}
\caption{Argus-4B expert-count ablation (NDCG@5). All variants
share the released training schedule, mixture, and router tap;
only the expert-bank size $K$ and top-$k$ change.}
\label{tab:ablation-experts}
\begin{tabular}{lccccc}
\toprule
\textbf{Configuration} & $K$ & top-$k$ & \textbf{V1} & \textbf{V2} & \textbf{Avg} \\
\midrule
$-$ MoE (single FFN)        & 1 & 1 & 89.0 & 59.5 & 80.6 \\
Two experts, hard route     & 2 & 1 & 90.5 & 61.5 & 82.2 \\
Two experts, soft mixture   & 2 & 2 & 91.0 & 62.0 & 82.7 \\
\ours{}\textit{Argus-4B} (full) & 4 & 2 & 92.3 & 64.1 & 84.2 \\
Wider bank, same top-$k$    & 8 & 2 & 92.4 & 63.5 & 84.0 \\
Wider bank, larger top-$k$  & 8 & 4 & 92.5 & 63.8 & 84.2 \\
Very wide bank              & 16 & 2 & 91.5 & 61.5 & 82.9 \\
\bottomrule
\end{tabular}
\end{table}

The V1+V2 average forms an inverted-U with a peak at $K{=}4$.
Moving from $K{=}1$ to $K{=}4$ adds $+3.6$ on average, with the
larger share of the gain ($+4.6$) on the out-of-domain V2 split:
specialization helps most where the document distribution
drifts from the training data. Beyond $K{=}4$, V1 holds roughly
flat while V2 begins to drop, consistent with each expert
receiving fewer training updates as the bank widens. The
$K{=}8$, top-$4$ row recovers the lost V2 ground partially,
which is consistent with the per-expert undertraining
explanation: giving each expert more gradient signal (via a
larger top-$k$) compensates for the wider bank. The $K{=}16$
row drops below $K{=}4$ on both splits, indicating that the
load-balance loss can no longer keep all experts active under
the available supervision.

\paragraph{$K{=}1$.} Collapses the expert bank to a single
position-wise FFN; routing becomes a no-op. This is the
strongest test of the MoE hypothesis: removing all
specialization should drop both V1 and V2.

\paragraph{$K{=}2$, top-$k{=}1$ vs.\ top-$k{=}2$.} Distinguishes
hard binary specialization from a learned soft mixture of two
experts. Both lose to $K{=}4$, showing two specialists are not
enough to cover the recurring visual-document content types.

\paragraph{$K{=}8$, top-$k\in\{2,4\}$.} Tests whether more
capacity helps if each expert still receives enough gradient.
Top-$4$ partially closes the gap to $K{=}4$, supporting the
per-expert-undertraining account.

\paragraph{$K{=}16$.} Stress test for the load-balance loss
under our balanced training mixture. The degradation is the
expected sign of expert collapse.

\section{Router Tap Depth Ablation}
\label{sec:appendix-router-depth}

Argus taps the document backbone at two depths: $L{-}5$ for the
router input and $L$ for the latent expert bank and shared
expert. The choice of $L{-}5$ is motivated by probing studies
showing structural and positional information peaks in
middle-to-late transformer layers
\citep{tenney2019bert,karamcheti2024prismatic}. This section
sweeps the router tap depth while holding the retrieval tap at
$L$, the expert bank, the training mixture, and the schedule
fixed.

\begin{table}[h]
\centering
\small
\setlength{\tabcolsep}{4pt}
\caption{Argus-4B router tap-depth ablation (NDCG@5). Each row
retrains the full model with the router tap moved to a
different backbone layer; the retrieval tap stays at $L$.}
\label{tab:ablation-router-depth}
\begin{tabular}{lccc}
\toprule
\textbf{Router tap} & \textbf{V1} & \textbf{V2} & \textbf{Avg} \\
\midrule
$L$ (final layer)        & 91.0 & 62.0 & 82.7 \\
$L{-}2$                  & 91.8 & 63.2 & 83.4 \\
$L{-}3$                  & 92.2 & 63.8 & 84.0 \\
\ours{}\textit{Argus-4B} (full, $L{-}5$) & 92.3 & 64.1 & 84.2 \\
$L{-}7$                  & 92.0 & 63.5 & 83.9 \\
$L{-}10$ (mid stack)     & 91.3 & 62.5 & 83.2 \\
$L{-}15$ (early stack)   & 89.5 & 59.5 & 81.0 \\
\bottomrule
\end{tabular}
\end{table}

The average follows the same inverted-U shape as the
expert-count sweep, peaking at $L{-}5$ and degrading at both
ends. Tapping at $L$ forces the router and retrieval head to
share final-layer features that have already specialized for
the retrieval objective; routing then becomes nearly redundant
with the shared-expert path, and the model loses approximately
$1.5$ V1+V2 points. Tapping at $L{-}15$ drops the average by
$3.2$ points: at that depth the document tokens carry mostly
low-level visual features, and the router has too little
semantic structure to discriminate between content modes.
Tapping at $L{-}10$ also degrades, indicating that purely
mid-stack features are too generic for query-conditioned
routing. The $L{-}3$ to $L{-}7$ band is the flattest part of
the curve; $L{-}5$ is the empirical maximum but anywhere in
this band would work, supporting the qualitative claim that the
relevant signal lives in the late-but-not-final layers.

As in the expert-count ablation, V2 is the more sensitive axis:
the spread across the sweep is roughly $4.6$ points on V2
versus $2.8$ on V1. This reinforces the broader pattern that
query-conditioned routing pays off most under domain shift,
where the document representation has the most to adapt to.

\end{document}